\documentclass[magazine, 10pt]{IEEEtran}
\usepackage{authblk}
\usepackage{amsmath}

\usepackage{amssymb,amsfonts}
\usepackage{algorithmic}
\usepackage{graphicx}
\usepackage{textcomp}
\usepackage{xcolor}
\def\BibTeX{{\rm B\kern-.05em{\sc i\kern-.025em b}\kern-.08em
    T\kern-.1667em\lower.7ex\hbox{E}\kern-.125emX}}
\usepackage[utf8]{inputenc}
\usepackage{booktabs} % For formal tables
\usepackage{graphicx}
\usepackage{subfigure}
\usepackage{units}
\usepackage{blindtext}
\usepackage{multirow}
\usepackage[sort, numbers]{natbib}
\usepackage{xcolor,xspace}
\usepackage{amsmath}
\usepackage{paralist}
\usepackage{mdframed}
\usepackage{pict2e}
\usepackage{mdframed}
\usepackage{listings}
\usepackage{color}
\usepackage{threeparttable}
\usepackage{array}
\usepackage{booktabs}
\usepackage{pifont}
\usepackage{balance}
\usepackage{url}
\usepackage{caption}

\usepackage{hhline}
\usepackage{footnote}
\usepackage{logicproof}
\usepackage{pbox}
\usepackage{lscape}

\usepackage{arydshln}
\usepackage{mdframed}
\usepackage{pgf-umlsd}

\usepackage[T1]{fontenc}
\usepackage{
  beramono,
  listings,
  textcomp
}

\usepackage{lipsum}

\usepackage[ruled]{algorithm2e}
\usepackage{algorithmic}
\LinesNumbered

\usepackage{tikz}

\usepackage{url}

\usepackage[
    n,
    advantage,
    operators,
    sets,
    adversary,
    landau,
    probability,
    notions,
    logic,
    ff,
    mm,
    primitives,
    events,
    complexity,
    asymptotics,
    keys]{cryptocode}

\usetikzlibrary{arrows, patterns, automata}
\usetikzlibrary{calc}
\usetikzlibrary{decorations.pathreplacing, shapes.multipart, shapes.geometric}

\usetikzlibrary{arrows,automata}
\usetikzlibrary{calc}
\usetikzlibrary{arrows,calc,shapes,decorations.pathreplacing}

\newcommand{\ignore}[1]{}

\newcommand{\acro}{{{\it $RATA$}}\xspace}
\newcommand{\acroa}{{{\it {$RATA_{A}$}}}\xspace}
\newcommand{\acrob}{{{\it {$RATA_{B}$}}}\xspace}

\newcommand{\vrased}{{{\sf\it VRASED}}\xspace}
\newcommand{\apex}{{{\sf\it APEX}}\xspace}
\newcommand{\pox}{{{\sf\it PoX}}\xspace}

\newcommand{\dev}{{\it Prover}\xspace}
\newcommand{\prv}{{\it Prover}\xspace}
\newcommand{\vrf}{{\it Verifier}\xspace}

\newcommand{\RA}{{\ensuremath{\sf{\mathcal RA}}}\xspace}

\definecolor{pigment}{rgb}{0.0, 0.65, 0.31}
\newcommand{\cmark}{{\large~\textcolor{pigment}{\ding{51}}}}%
\newcommand{\xmark}{{\large~\textcolor{red}{\ding{55}}}}%

\newcommand{\smart}{SMART\xspace}

\newcommand{\attkey}{\ensuremath{\mathcal K}\xspace}

\renewcommand\adv{\ensuremath{\sf{\mathcal Adv}}\xspace}

\mathchardef\mhyphen="2D

\newcommand{\toctou}{{\small TOCTOU}\xspace}

\newcommand{\sw}{\texttt{\small SW-Att}\xspace}

\newcommand{\CFA}{{\ensuremath{\sf{\mathcal CFA}}}\xspace}
\newcommand{\tinycfa}{{{\sf\it Tiny-CFA}}\xspace}
\newcommand{\cflog}{{\small CF-Log}\xspace}

\newcommand{\DFA}{{\ensuremath{\sf{\mathcal DFA}}}\xspace}
\newcommand{\dialed}{{{\sf\it DIALED}}\xspace}
\newcommand{\ilog}{{\small I-Log}\xspace}

\makeatletter
\def\ps@IEEEtitlepagestyle{%
  \def\@oddfoot{\mycopyrightnotice}%
  \def\@oddhead{\hbox{}\@IEEEheaderstyle\leftmark\hfil\thepage}\relax
  \def\@evenhead{\@IEEEheaderstyle\thepage\hfil\leftmark\hbox{}}\relax
  \def\@evenfoot{}%
}
\def\mycopyrightnotice{%
  \begin{minipage}{\textwidth}
  \centering \scriptsize
  Copyright~\copyright~20xx IEEE. Personal use of this material is permitted. Permission from IEEE must be obtained for all other uses, in any current or future media, including reprinting/republishing this material for advertising or promotional purposes, creating new collective works, for resale or redistribution to servers or lists, or reuse of any copyrighted component of this work in other works.
  \end{minipage}
}
\makeatother
\author{
\IEEEauthorblockN{%%
   Ivan De Oliveira Nunes\IEEEauthorrefmark{1},
   Sashidhar Jakkamsetti\IEEEauthorrefmark{2},
   Norrathep Rattanavipanon\IEEEauthorrefmark{3},
   Gene Tsudik\IEEEauthorrefmark{4}\\
}
\IEEEauthorblockA{%%
  \IEEEauthorrefmark{1}Rochester Institute of Technology, USA\\
  \IEEEauthorrefmark{2}Robert Bosch LLC, USA\\
  \IEEEauthorrefmark{3}Prince Songkla University -- Phuket Campus, Thailand\\
  \IEEEauthorrefmark{4}University of California Irvine, USA\\
}
}

\usepackage[normalem]{ulem} 
\newcommand\add[1]{#1}
\newcommand{\remove}[1]{}

\begin{document}
	
\title{Towards Remotely Verifiable Software Integrity in Resource-Constrained IoT Devices}
\maketitle

\begin{abstract}
Lower-end IoT devices typically have strict cost constraints that rule out usual security mechanisms 
available in general-purpose computers or higher-end devices.  To secure low-end devices, various 
low-cost security architectures have been proposed for remote verification of their software state 
via integrity proofs. These proofs vary in terms of expressiveness, with simpler ones confirming 
correct binary presence, while more expressive ones support verification of arbitrary code execution.
This article provides a holistic and systematic treatment of this family of architectures. It also 
compares (qualitatively and quantitatively) the types of software integrity proofs, respective 
architectural support, and associated costs. Finally, we outline some research directions and emerging
challenges.
\end{abstract}

\section{Introduction}
    Micro-Controller Units (MCUs) that perform actuation and/or sensing are the \textit{de facto} interfaces 
    between the analog and digital worlds. On actuators, digital commands are converted into physical actions, 
    while sensors convert analog ambient quantities into digital form.  They represent the point where 
    data is ``born'' and first processed. At the same time, since MCUs are programmable, their software 
    can be compromised and subsequently corrupt data, e.g., by modifying software to forge/spoof a 
    sensed values or ``lie'' about having performed actuation commands. One na\"ive defense strategy is
    to make all software non-writable, e.g. by housing it in ROM. While this obviates software compromise, 
    it also precludes legitimate software updates.

    In the last decade, the research community has actively identified and examined this issue~\cite{RA2022survey}. 
    Earlier results proposed methods to allow a trusted party called \vrf to remotely check if the correct 
    binary is currently installed on a remote and untrusted \dev. This security service is 
    commonly known as \emph{Remote Attestation (\RA)}~\cite{smart,vrasedp,pistis,trustlite}.
    
    A related notion is Proofs of Execution (\pox)~\cite{apex} which extends \RA to prove correct execution 
    of the attested binary or parts thereof, i.e., functions within the binary. In line with \pox, Control Flow 
    Attestation (\CFA) allows \vrf to also verify the sequence of executed instructions. This detects 
    software exploits that corrupt execution path by changing the program control 
    flow without modifying the actual binary (aka code-reuse 
    attacks~\cite{eternalwar}). Data-Flow Attestation (\DFA) further extends \CFA with detection of 
    data-only attacks that exploit vulnerabilities to corrupt data without modifying the program control flow.
	
	This article overviews a series of recent low-cost techniques, based on HW/SW co-design, that create unforgeable proofs of software integrity (encompassing aforementioned services) for the MCUs 
    commonly used in low-end IoT devices. Each of these technique tackles one of the following questions:
    \begin{compactenum}
     \item How to prove that an MCU of interest is currently installed with the correct software/firmware binary?
     \item How to extend this proof with historical context, i.e., how to determine ``since when'' the expected software has been installed on the device?
     \item Upon receiving a result from a remote MCU (e.g., a sensed value), how to ensure that it was indeed obtained through the proper execution of expected software on the expected device?
     \item Can we verify that instructions were executed in the intended/legal order? 
     In other words, how to ensure the absence of control flow attacks during the execution?
     \item In addition, how to ensure the absence of (non-control) data-only attacks during execution?
    \end{compactenum}
    Naturally, approaches that provide more expressive evidence (thereby detecting stealthier attacks) 
    also incur higher hardware and run-time overhead.
    
    Folklorically and historically, the common wisdom holds that the typical/usual low-end MCUs (such as 
    TI MSP430
    %\footnote{https://www.ti.com/microcontrollers-mcus-processors/microcontrollers/msp430-microcontrollers/overview.html} 
    or AVR ATMega)
    %\footnote{https://www.microchip.com/en-us/product/ATmega32}
    are incapable of supporting software integrity verification. While this is true for unmodified MCUs, recent 
    research results show that it can indeed be achieved with
    minimal hardware overhead, low overall cost, and strong security guarantees.  
    This article overviews and compares (qualitatively and quantitatively) a sequence of 
    five techniques, each incrementally addressing one of the above questions. We also identify several 
    outstanding challenges that need to be tackled by future work.    

\section{Background}
\subsection{Resource-Constrained/Low-End MCUs}\label{sec:scope}
\begin{figure}
	\centering
    \includegraphics[width=.8\columnwidth]{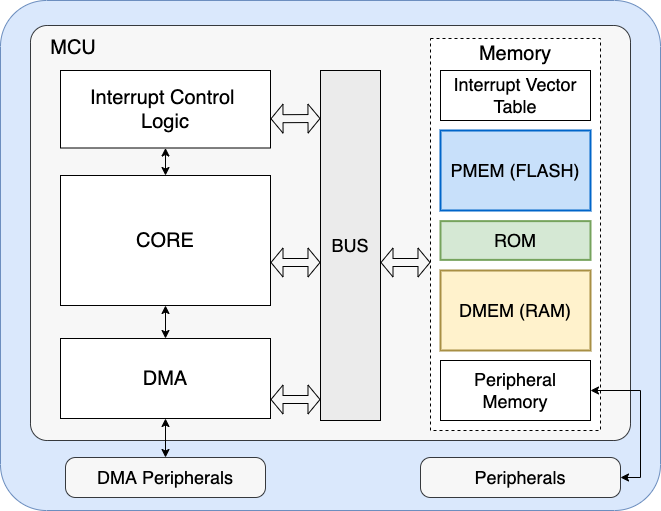}
    \caption{Architecture of Low-end IoT Devices}
    \label{fig:mcu}
\end{figure}
This article focuses on resource-constrained embedded/smart/IoT sensors and actuators (or hybrids thereof). 
These are some of the simplest and smallest computing devices, based on low-power 
single-core MCUs with only a few KBytes of memory. 
Figure \ref{fig:mcu} illustrates a typical MCU architecture, featuring a CPU core, an Interrupt Control Logic module, and a Direct Memory Access (DMA) controller connected to main memory via a bus.
The MCU includes four memory types: (1) program memory (PMEM), (2) read-only memory (ROM), (3) data memory (DMEM), and (4) peripheral memory. Application software is stored in PMEM, usually realized as non-volatile physical memory such as Flash or FRAM. Runtime data is stored in volatile DMEM, implemented using RAM. ROM contains the bootloader and any software fixed at the time of manufacturing or provisioning, and remains immutable thereafter. DMA can read and write memory in parallel with the core. 
Generally, low-end MCUs run software atop ``bare metal'', i.e., execute software directly from PMEM, without relying on memory management units (MMU) and often not even memory protection units (MPU). Examples of such MCUs include Atmel AVR ATmega, TI MSP430, %\add{and ARM Cortex-M}, featuring $8$- and $16$-bit single-core CPUs, running at clock frequencies of $1$-$16$MHz, with $64$ KB of addressable memory split among ROM, DMEM, and PMEM.
% Sashi: If we are including Cortex-M, we might have to modify other numbers like below?
\add{and ARM Cortex-M, featuring $8/16/32$-bit single-core CPUs, running at clock frequencies of $1$-$48$MHz, with up to $128$ KB of addressable memory.}

\subsection{Attack Vectors \& Threat Model}
To reason about software integrity one must consider that all modifiable memory could be 
tampered with by the adversary (\adv), unless explicitly protected by the hardware architecture. Therefore,  
\adv is assumed to control the entire software state of \prv.
This allows \adv to read and write any memory that is not explicitly protected by hardware, program DMA controllers, and trigger interrupts at any given moment.

%\adv can modify any writable memory and read any memory (including secrets) that is not explicitly protected by hardware. Also, \adv can program DMA controllers as well as trigger any interrupts at any point in time. 

While \adv may reprogram \prv software in PMEM via a wired interface (e.g., USB or J-TAG), we consider invasive attacks physically altering hardware and ROM code to be out of scope. 
%Specifically, \adv cannot: (1) modify hardware components, (2) alter ROM code, (3) cause physical hardware faults, or (4) extract secrets via physical side-channels. 
Protection against these attacks can be obtained through orthogonal tamper-resistance techniques, \add{e.g., employing internal power regulators, implementing anomaly detection for the MCU's behavior, enclosing the MCU with additional metal layers or enforcing physical access control to the devices.}

\ignore{
\begin{figure}
	\centering
	\resizebox{0.6\columnwidth}{!}{%
		\includegraphics[]{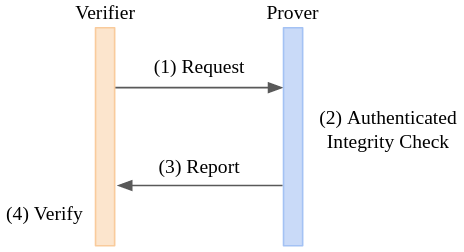}
	}
	\caption{Typical \RA interaction between \vrf and \prv}
	\label{fig:RA}
\end{figure}
}

\section{Verifying Software/Firmware Integrity with \RA}\label{sec:vrased}

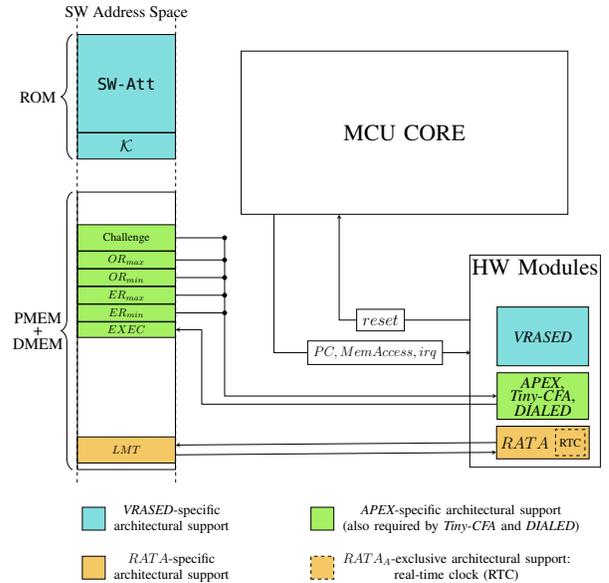
\begin{figure}[!ht]
	\centering
	\resizebox{.9\columnwidth}{!}{%
		\begin{tikzpicture}[node distance=1.5cm, >=stealth]
\tikzset{
	msp430/.style={
		draw,
		rectangle,
		minimum height=5cm,
		minimum width=10cm,
		align=center
	},
	membb/.style={
		draw,
		rectangle,
		minimum height=9cm,
		minimum width=1.5cm,
		align=center
	},
	mem/.style={
		draw,
		rectangle,
		minimum height=.5cm,
		minimum width=1.5cm,
		align=center
	},
	smallbox/.style={
		draw,
		rectangle,
		minimum height=1.8cm,
		minimum width=2.8cm,
		align=center
	},
	hwmod/.style={
		draw,
		rectangle,
		minimum height=6.5cm,
		minimum width=4cm,
		align=center
	},
	tinybox/.style={
		draw,
		rectangle,
		minimum height=.2cm,
		minimum width=3cm,
		align=center
	},
	chalbox/.style={
		draw,
		rectangle,
		minimum height=.8cm,
		minimum width=3cm,
		align=center
	},
	swattbox/.style={
		draw,
		rectangle,
		minimum height=3cm,
		minimum width=3cm,
		align=center
	},
	keybox/.style={
		draw,
		rectangle,
		minimum height=.8cm,
		minimum width=3cm,
		align=center
	},
	erbox/.style={
		draw,
		rectangle,
		minimum height=4cm,
		minimum width=3cm,
		align=center
	},
	hwsmallbox/.style={
		draw,
		rectangle,
		minimum height=1cm,
		minimum width=2.8cm,
		align=center
	},
	ratabox/.style={
		draw,
		rectangle,
		minimum height=1cm,
		text width=2.6cm,
		align=left
	},
	orbox/.style={
		draw,
		rectangle,
		minimum height=1cm,
		minimum width=3cm,
		align=center
	},
	membox/.style={
		draw,
		rectangle,
		minimum height=8.5cm,
		minimum width=3cm,
		align=center
	},
	lmtbox/.style={
		draw,
		rectangle,
		minimum height=.8cm,
		minimum width=3cm,
		align=center
	},
        rataa/.style={
		draw,
		rectangle,
		minimum height=.8cm,
            dashed,
            thick
        }
}
\definecolor{vrasedcolor}{RGB}{130,222,222}
\definecolor{apexcolor}{RGB}{166,240,100}
\definecolor{ratacolor}{RGB}{244,200,100}
\node[msp430] (msp) at (0,0) {\huge MCU CORE};
\node[smallbox, fill=vrasedcolor, below right = 4cm of msp.south] (vrased) {\Large \vrased};
\node[hwsmallbox, fill=apexcolor, below = .2cm of vrased.south] (vape) {\Large \apex, \\ \Large \tinycfa, \\ \Large \dialed};
\node[ratabox, fill=ratacolor, below = .2cm of vape.south] (rata) {\Large \acro};
\node[rataa, fill=ratacolor, right = 1.8cm of rata.west] (rataa) {RTC};
\node[hwmod] at (4,-7) (hwmod) {};
\node[overlay,anchor=north] at (hwmod.north) (test) {\huge HW Modules};

\path (msp.south) +(-2,0) coordinate (msp_0);
\path (msp.south) +(-4,0) coordinate (msp_1);
\path (vrased.west) +(-.8,.5) coordinate (vrased_0);
\path (vrased.west) +(-.8,-.5) coordinate (vrased_1);
\draw[thick, ->] (msp_1) |- (vrased_1); 
\draw[thick, <-] (msp_0) |- (vrased_0) ; 
\node[inner sep=5pt,right, fill=white, draw] at ($(vrased_0) + (-3.5,0)$) {\Large $reset$};

\node[inner sep=5pt,right, fill=white, draw] at ($(vrased_1) + (-5,0)$) {\large $PC, Mem Access, irq$};

% ROM
\node[swattbox, fill=vrasedcolor, above left = 0cm and 2cm of msp.west] (swatt) {\texttt{\Large SW-Att}};
\node[keybox, fill=vrasedcolor, below = 0cm of swatt.south] (key) {\Large \attkey};

\draw [decorate,decoration={brace,amplitude=10pt,raise=4pt},yshift=0pt] 
(key.south west) -- (swatt.north west) node [black,midway,xshift=-1.2cm] {\Large ROM};

% METADATA
\node[chalbox, fill=apexcolor, below = 2cm of key.south] (chal) {Challenge};
\node[tinybox, fill=apexcolor, below = 0cm of chal.south] (ormax) {$OR_{max}$};
\node[tinybox, fill=apexcolor, below = 0cm of ormax.south] (ormin) {$OR_{min}$};
\node[tinybox, fill=apexcolor, below = 0cm of ormin.south] (ermax) {$ER_{max}$};
\node[tinybox, fill=apexcolor, below = 0cm of ermax.south] (ermin) {$ER_{min}$};
\node[tinybox, fill=apexcolor, below = 0cm of ermin.south] (exec) {$EXEC$};

% ER
%\node[erbox, below = 1cm of exec.south] (er) {$ER$};
% OR
%\node[orbox, below = .5cm of er.south] (or) {$OR$};

% Arrow from APEX to EXEC
\path (vape.west) +(0,0) coordinate (vape_top);
\path (vape.west) +(0,-.25) coordinate (vape_bot);
\path (exec.east) +(.8,0) coordinate (exec_right);
\draw[thick, -] (vape_bot) -| (exec_right); 
\draw[thick, ->] (exec_right) -- (exec.east); 

% Arrow from APEX to \chal, ormin/max, ermin/max
\path (chal.east) +(1.5,0) coordinate (chal_right);
\path (ormin.east) +(1.5,0) coordinate (ormin_right);
\path (ormax.east) +(1.5,0) coordinate (ormax_right);
\path (ermin.east) +(1.5,0) coordinate (ermin_right);
\path (ermax.east) +(1.5,0) coordinate (ermax_right);
\draw[thick, <-] (vape_top) -| (chal_right); 
\draw[thick, -] (chal_right) -- (chal.east); 
\draw[thick, -] (ormin_right) -- (ormin.east); 
\draw[thick, -] (ormax_right) -- (ormax.east); 
\draw[thick, -] (ermin_right) -- (ermin.east); 
\draw[thick, -] (ermax_right) -- (ermax.east); 

% Draw line to connect vrased's input to apex
%\path (vrased_1) +(-.5, 0) coordinate (vrased_dot);
%\path (vape.west) +(0,0.25) coordinate (vape_connect);
%\draw[thick, ->] (vrased_dot) |- (vape_connect); 

% Draw line to connect vrased's input to rata
%\path (vrased_1) +(-.5, 0) coordinate (vrased_dot);
%\path (rata.west) +(0,0.25) coordinate (rata_connect);
%\draw[thick, ->] (vrased_dot) |- (rata_connect); 

%\path (vrased_dot) +(0,-.85) coordinate (dot1);
%\node at (dot1)[circle,fill,inner sep=1.5pt]{};
%\path (vrased_dot) +(0,-2.05) coordinate (dot2);
%\node at (dot2)[circle,fill,inner sep=1.5pt]{};

% Add dots
\foreach \n in {chal_right, ormin_right, ormax_right, ermin_right, ermax_right}%, vrased_dot}
\node at (\n)[circle,fill,inner sep=1.5pt]{};

% Memory
\node[membox, below = 1cm of key.south] (mem) {};

% LMT
\node[lmtbox, fill=ratacolor, above = .2cm of mem.south] (lmt) {$LMT$};

% Arrow from RATA to LMT
\path (rata.west) +(0,0) coordinate (rata_top);
\path (rata.west) +(0,-.3) coordinate (rata_bot);
\path (lmt.east) +(0,-.15) coordinate (lmt_bot);
\path (lmt.east) +(0,.15) coordinate (lmt_top);
\draw[thick, ->] (rata_top) -- (lmt_top); 
\draw[thick, ->] (lmt_bot) -- (rata_bot); 

\draw [decorate,decoration={brace,amplitude=10pt,raise=4pt},yshift=0pt] 
(mem.south west) -- (mem.north west) node [black,midway,xshift=-1.2cm] {\Large\shortstack{PMEM \\ + \\ DMEM}};

% Draw dashed line to indicate memory
\path (swatt.north west) +(0,0.5) coordinate (memleft_start);
\path (mem.south west) +(0,-0.5) coordinate (memleft_end);
\draw[dashed, thick, -] (memleft_start) -- (memleft_end); 
\path (swatt.north east) +(0,0.5) coordinate (memright_start);
\path (mem.south east) +(0,-0.5) coordinate (memright_end);
\draw[dashed, thick, -] (memright_start) -- (memright_end); 

% Say it's memory
\path (swatt.north) +(0, 1) coordinate (mem_name);
\node[overlay,anchor=north] at (mem_name) (test) {\Large SW Address Space};

%\node at ($(exec)!.5!(lmt)$) {\Large\textbf{\shortstack{.\\ \\.\\ \\.}}};

% Legend
\path (mem.south) +(-1,-1.5) coordinate (vrased_legend);
\node at (vrased_legend) [draw, rectangle, fill=vrasedcolor, minimum height=.7cm,minimum width=.7cm] {};
\path (vrased_legend) +(.5,0) coordinate (vrased_legend_text);
\node[overlay,anchor=west] at (vrased_legend_text) (testt) {\large \shortstack{\vrased-specific\\ architectural support}};

\path (mem.south) +(6,-1.5) coordinate (apex_legend);
\node at (apex_legend) [draw, rectangle, fill=apexcolor, minimum height=.7cm,minimum width=.7cm] {};
\path (apex_legend) +(.5,0) coordinate (apex_legend_text);
\node[overlay,anchor=west] at (apex_legend_text) (testt) {\large \shortstack{\apex-specific architectural support\\ (also required by \tinycfa and \dialed)}};

\path (mem.south) +(-1,-3) coordinate (rata_legend);
\node at (rata_legend) [draw, rectangle, fill=ratacolor, minimum height=.7cm,minimum width=.7cm] {};
\path (rata_legend) +(.5,0) coordinate (rata_legend_text);
\node[overlay,anchor=west] at (rata_legend_text) (testt) {\large \shortstack{\acro -specific\\ architectural support}};

\path (mem.south) +(6,-3) coordinate (rataa_legend);
\node at (rataa_legend) [draw, dashed, rectangle, fill=ratacolor, minimum height=.7cm,minimum width=.7cm] {};
\path (rataa_legend) +(.5,0) coordinate (rataa_legend_text);
\node[overlay,anchor=west] at (rataa_legend_text) (testt) {\large \shortstack{\acroa-exclusive architectural support:\\ real-time clock (RTC)}};

\end{tikzpicture}
	}
	\caption{Architectural requirements of various integrity proofs}
	\label{fig:arch}
\end{figure}

Completely preventing illegal code modifications in low-end devices is a challenging task
due to the need to perform code updates. As an alternative, \RA is an inexpensive and 
effective technique which detects attacks that modify \prv code.
\RA allows \vrf to remotely assess software integrity of \prv, in an on-demand fashion.
%As shown in Figure~\ref{fig:RA}, 
This is typically realized as a \vrf-initiated 
challenge-response protocol where:
\begin{compactenum}
    \item \vrf sends an attestation request containing a cryptographic challenge to \prv.
    \item \dev performs an {\em authenticated integrity check} based on the received challenge over its own PMEM.
    \item \dev returns the result to \vrf.
    \item \vrf validates whether the result matches a valid PMEM state by comparing it to the expected (benign) value. 
\end{compactenum}
The purpose of the challenge in step 1 is to ensure that \prv's response is fresh, i.e., reflects 
its current software state (rather than an old replayed response).
The {\em authenticated integrity check} in step 2 can be implemented as a Message Authentication Code (MAC) 
or signature. We refer to the function that implements the authenticated integrity check as the {\em integrity-ensuring function (IEF)}.
To implement IEF, \prv must maintain a secret key (\attkey), confidentiality of which 
must be preserved even if \prv software is compromised.
Consequently, main challenges in designing secure \RA revolve around: (i) secure storage of \attkey, 
and (ii) establishment of an immutable secure run-time environment that accesses \attkey to compute 
the IEF without leaking \attkey to any other software in the \prv.

SANCUS~\cite{Sancus17} protects the IEF and \attkey 
by implementing them entirely in hardware inaccessible to untrusted software. This eliminates the need 
for any software component of the trusted computing base (TCB). However, this approach incurs a 
significant hardware cost, which may be prohibitive for budget-coscious low-end MCUs.  
\smart~\cite{smart} is designed to minimize hardware overhead by using a hybrid (HW/SW co-design) \RA approach.
\smart implements its IEF in software, while a small amount of hardware is used
to detect any violation that attempts to leak \attkey or tamper with IEF execution. 
The main trade-off between SMART and SANCUS is speed {\em vs} cost -- being all
hardware, the latter is faster, while the former is cheaper.
	
Building upon SMART design principles in a less {\em ad hoc} fashion, recently proposed 
\vrased~\cite{vrasedp} technique is a formally verified hardware/software \RA co-design.
Figure~\ref{fig:arch} shows its hardware and software components. 
The trusted software module (\sw) contains IEF code and \attkey stored in ROM.
This way, neither \sw nor \attkey values can be modified after manufacturing or provisioning.
\vrased also includes a hardware monitor that tracks several MCU signals to determine: 
(1) $PC$ value, i.e., address of currently executing instruction; (2) $MemAccess$, memory address 
currently being read or written by either MCU Core or DMA; and (3) $irq$, a one-bit 
signal indicating whether an interrupt is currently being triggered.

Using these signals, \vrased hardware monitor detects violations that try to violate
secrecy of \attkey or \sw execution integrity through:
\begin{compactenum}
\item Illegal accesses to \attkey by any software other than \sw. 
% since access to any trace of \attkey that may exist in the MCU's DMEM at runtime.
\item Any incomplete or interrupted \sw execution that could lead to forgery of an attestation result. 
\end{compactenum}
Upon detecting a violation, \vrased triggers an immediate MCU $reset$, promptly preventing the violation.

To facilitate formal verification, \vrased avoids state explosion problems by structuring its 
implementation as a collection of sub-modules, each guaranteeing a specific set of formal sub-properties.
Each sub-module undergoes individual verification, and the combination of all sub-modules is then verified for end-to-end notions of \RA soundness and security.
Informally, {\bf \RA soundness} ensures correct IEF computation over current PMEM, while 
{\bf \RA security} guarantees that IEF execution 
produces an unforgeable authenticated PMEM measurement and prevents \attkey leakage before, during, or after \RA. 
Further details on \vrased formal verification can be found in \cite{vrasedp}.
	
%\vrased hardware is described at Register Transfer Level (RTL) using Finite State Machines (FSMs). Next, 
%a model checker is used to automatically prove that FSMs achieve claimed security sub-properties. 
%Finally, the proof that the conjunction of hardware and software sub-properties implies end-to-end 
%soundness and security is performed using an LTL theorem prover. 
%We refer to \cite{vrasedp} for further details.
	
\section{TOCTOU-Security \& Efficient \RA}
Recall that each \RA instance is initiated by \vrf. An authentic \RA result received from 
\prv reflects \prv code \textit{only at the time when it is computed}.
In particular, it provides no information about \prv software \textbf{before} \RA execution 
or \textbf{between} successive \RA instances. 
This issue is commonly termed as \emph{Time-Of-Check Time-Of-Use} or \toctou. In the context of \RA, \toctou
means that transient malware can not be detected by \RA. Concretely, if malware infects \prv, performs its 
malicious tasks, and erases itself prior to the next \RA instance, its ephemeral presence would remain 
unnoticed.
	
\acro~\cite{rata} is another recent technique to address the \toctou problem. It extends \vrased  
with a minimal and formally verified hardware component that additionally provides historical context 
about the state of PMEM. \acro consists of two alternative designs. Shown as yellow components 
in Figure~\ref{fig:arch}, the first version -- \acroa\ -- is a verified hardware module that operates 
as follows:
\begin{compactenum}
    \item It monitors MCU $PC$ and $MemAccess$ signals and uses this information to detect whether PMEM is 
    currently being modified. 
    \item When a PMEM modification is detected, \acroa retrieves the current time from a real-time clock (RTC) and stores it in a designated and secure memory area, called the Latest Modification Time ($LMT$) region
    \item $LMT$ region is always covered by IEF  (i.e., included in each attestation result, along with PMEM) and read-only to all software and DMA.
    %This means the time of the latest PMEM modification can be inferred from the attestation result. 
    %$LMT$ is also read-only to all software (and DMA). This is enforced by \acroa verified hardware. 
\end{compactenum}
To verify the attestation result, \vrf compares the received $LMT$ value with the time of the last authorized PMEM 
modification (usually, time of latest legitimate code update) to check whether any unauthorized 
activity occurred since then.
	
In practice, RTCs are usually unavailable on resource-constrained devices and secure clock synchronization 
in distributed systems poses a significant challenge, particularly for such devices. \acroa is thus useful only 
to demonstrate the general approach. To make it practical, the second version (\acrob) eliminates the RTC 
requirement. \acrob relies on \vrf's own notion of time by associating each attestation challenge 
to the time of its issuance by \vrf.
\prv logs the latest received challenge to $LMT$. For this to work, each \vrf's challenge must be unique for each 
\RA instance, to prevent replay attacks. This allows \acrob to uniquely
associate each challenge to \vrf's notion of time.
\acrob hardware component is similar \acroa, except that $LMT$ is now updated with the current
challenge {\bf if and only if} a PMEM modification occurred since the previous \RA instance.

An important side benefit of \acro is its ability to significantly reduce \RA execution time on \prv since it is no longer necessary to compute the IEF over the entire PMEM most of the time.
Assuming that \vrf already knows PMEM contents from a previous \RA result, it is adequate to demonstrate that no changes have occurred since then.
This can be achieved by attesting {\bf only $LMT$} as opposed to the entire PMEM, resulting in a 
substantial reduction of computation time. In fact, this time is constant in size of $LMT$, instead of 
increasing linearly with PMEM size. 
\remove{For instance, given a typical $8$ kBytes PMEM, \acro 
takes roughly $50$ms on \prv, as opposed to $1$sec in VRASED.}
\add{For instance, on an MSP430 MCU running at 8MHz with $8$ kBytes PMEM, \acro 
takes roughly $50$ms on \prv, as opposed to $1$sec in VRASED.}
(See~\cite{rata} for details).

\section{From \RA to Proofs of Execution}
\RA yields an indication of whether \prv PMEM contains expected code.
However, it does not guarantee that any operation (i.e., sensing/actuation functions in that code) was 
executed correctly. Also, it does not bind results (e.g., sensor readings) to the correct execution of appropriate
code. In other words, \RA gives no secure association between data received by \vrf from the \prv and \prv's execution of a specific application-dependent operation. 
Thus, \adv can tamper with or spoof data even if \prv PMEM contains correct code.

For this reason, \apex~\cite{apex} proposes the concept of ``Proofs of Execution'' (PoX) by augmenting 
\RA to prove to \vrf that:
\begin{compactenum}
\item The function of interest exists within a specific region of PMEM;
\item This function was indeed executed in a timely manner, upon \vrf's request; and
\item Any claimed output was indeed produced by this timely execution of the desired function on \prv.
\end{compactenum}
Hence, PoX enables authentication of data from its ``birth'', i.e., at the point when it
becomes digital, through the interaction of code with sensing ports (e.g., general purpose I/O).

\apex~\cite{apex} structure is shown in green in Figure~\ref{fig:arch}. It is built atop a secure 
\RA technique, such as \vrased, by introducing an additional hardware module that controls 
a $1$-bit flag, called $EXEC$, that can not be modified by any software. 
The key is to use the high value of $EXEC$ to inform \vrf that the intended portion of attested code 
(a \vrf-defined code section in PMEM) was executed successfully between the time when \vrf challenge was 
issued by \vrf and the time when IEF on \prv executed. 
Similarly, a value of 0 for $EXEC$ indicates that execution 
of that code section did not occur, or that it was tampered with.

\apex IEF covers: $EXEC$ flag itself; the region where output must be saved 
(called output region or $OR$); and code stored in a \vrf-defined section of PMEM 
(called executable region or $ER$). 
Security of the \RA architecture guarantees the contents of these memory regions (including $EXEC$) 
can not be spoofed. Therefore, as long as \apex hardware properly controls $EXEC$, the \RA result 
constitutes unforgeable proof that code in $ER$ was executed and produced the results stored in $OR$.
\apex considers that code section executed properly (setting $EXEC$ to $1$) if and only if:
\begin{compactenum}
    \item Execution of $ER$-resident code is atomic (i.e., uninterrupted), from $ER$'s first, to its last, instruction;
	
    \item Neither code in $ER$, nor its output in $OR$ is modified between execution and the next IEF computation;

 %%GTS: RAM??? Or PMEM?
 %OAK: RAM (to prevent tampering with immediate values leading to OR value's modification), PMEM is already covered in the previous one (ER immutability)
    \item During execution, DMEM is not be modified by DMA or by other software functions except $ER$.
\end{compactenum}
These conditions ensures that when $EXEC$ is 1, both $ER$ and $OR$ remain consistent between $ER$ 
code execution and subsequent IEF computation, plus execution itself is not tampered with. $ER$ and $OR$ 
locations and sizes are configured using values of $ER_{min}$, $ER_{max}$, $OR_{min}$ and $OR_{max}$ in DMEM, 
as shown in Figure~\ref{fig:arch}.
This allows \apex to support PoX of arbitrary code and output sizes.
%as long as these configurations are 
%performed before $ER$ execution. 
%To ensure temporal consistency, any attempt to modify these parameters and the attestation challenge after $ER$ execution starts causes \apex to set $EXEC$ to 0.

%%GTS: Which properties??? 
%OAK: changed to "verified w.r.t. the aforementioned security requirements"
\apex hardware is verified to conform to formal specifications of the abovementioned $EXEC$ behavior. 
These specifications, along with the underlying verified guarantees of \vrased, are proven to guarantee a security definition for unforgeable \pox, as discussed in~\cite{apex}.

\begin{figure*}
	\centering
    \includegraphics[width=\textwidth]{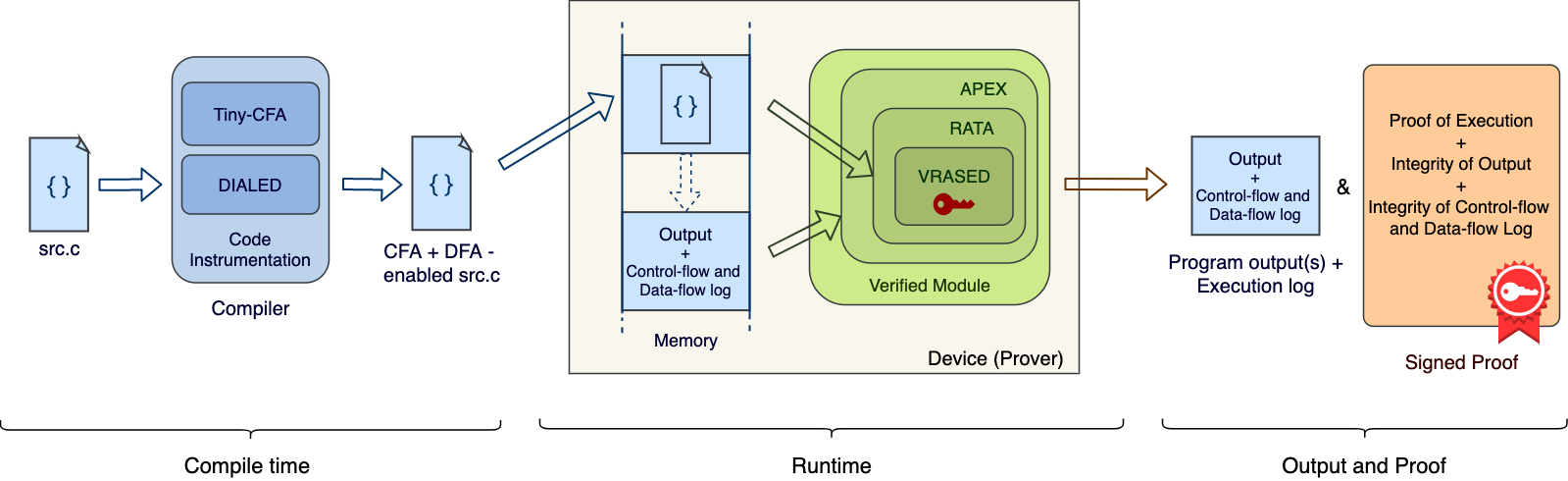}
    \caption{Phases in Embedded Software Execution Integrity}
    \label{fig:workflow}
\end{figure*}

\section{\CFA on Resource-Constrained MCUs: Augmenting PoX to Verify Control Flow Paths}
\pox assumes that the code for which execution is being proven (i.e., the code in $ER$) is free of memory-safety vulnerabilities, such as those leading to buffer overflows and similar attacks. 
However, when these vulnerabilities (unintentionally) exist in the executable, they can be exploited at the time of execution to launch well-known control flow attacks (such as return- and jump-oriented programming) that change the order in which the instructions are executed to cause unintended behavior, without modifying the program's code. As a consequence, these attacks would remain oblivious to \RA or \pox. 
	
Control Flow Attestation (\CFA) aims to detect control flow attacks by also providing \vrf with a report that shows the exact order in which the instructions that form a software operation of interest have executed on \prv. 
This can be accomplished by securely recording the destination of every control flow altering instruction, such as jumps, branches, and returns, during the program's execution.

A number of \CFA techniques have been proposed in recent years (e.g., C-FLAT~\cite{cflat} and LiteHAX~\cite{dessouky2018litehax}). However, they target higher-end embedded devices (e.g., those featuring application CPUs such as Raspberry Pi). Unfortunately, these techniques are prohibitively expensive for resource-constrained MCUs. 

\tinycfa\cite{tinycfa} was recently developed to address the \CFA problem in the context of resource-constrained MCUs by leveraging inexpensive \pox as its only hardware requirement.
As shown in Figure~\ref{fig:workflow}, \tinycfa introduces an additional compilation-time phase where the code to be executed by \prv is instrumented with additional instructions that generate a log (referred to as \cflog), containing the control flow path taken during execution.

During a \pox of the instrumented code, execution yields \cflog, in addition to its regular result/output. %, which is then included in the attestation result. 
\tinycfa ensures authentication and integrity of \cflog by making \cflog a part of the \pox output, which is located within \apex's output region $OR$ and covered by the IEF.
%, thus inheriting the execution's output integrity assurance from \apex.
As a result, \vrf can use this new evidence (\cflog) to determine the validity of the execution control flow path and verify the absence of control flow hijacking attacks.

In more detail, \tinycfa instruments the executable to ensure that \cflog contains all information required by \vrf to reconstruct the control flow path by:
	
\begin{itemize}
    \item {\it Securely logging control flow instructions:} all control flow altering instructions are prepended with additional instructions to log their destinations to \cflog. 
    
    \item {\it Ensuring append-only \cflog:} 
    %Instrumentation ensures that \cflog can never be overwritten. 
    direct writes to \cflog are replaced at compile time while indirect writes are instrumented to check whether their destination is within \cflog at runtime. Upon detecting an illegal write to \cflog, the \pox is halted, implying an invalid control flow. 
\end{itemize}

Due to the resource-constrained nature of MCUs, \CFA schemes should have minimal hardware and runtime overheads. \tinycfa minimizes hardware requirements by requiring no hardware support other than \pox from \apex. It also implements several optimizations to keep the runtime overhead and \cflog size within practical limits. We discuss these overheads in Section~\ref{sec:comparison} and revisit opportunities for future work on reducing \CFA runtime costs in Section~\ref{sec:opportunities}.
	
\section{MCU Data Flow Integrity atop \CFA}
	
Aside from control flow attacks (detected by \CFA), stealthier attacks known as ``data-only'' attacks can still originate from memory safety vulnerabilities. Specific vulnerabilities (see example in~\cite{dialed}) allow attacks to corrupt intermediate data variables in DMEM without even altering the control flow of the program
(hence ``data-only''). %here attacks because they primarily target to change the data variables of the program and not the control flow. Detection of such data-only attacks still remains elusive. 
%Without a way to detect data-only attacks (as well as control-flow modifications), the results of \prv's remote computation cannot be fully trusted.
	
Detection of such data-only attacks still remains elusive and requires verifying the data-flow integrity during execution -- a service known as Data-Flow Attestation (\DFA). Prior work in \DFA such as OAT~\cite{sun2020oat} requires user annotations and relatively expensive trusted hardware support. % (e.g., ARM TrustZone), which is not affordable to low-end MCUs. 

\dialed\cite{dialed} presents the first \DFA architecture aimed at resource-constrained MCUs by following an approach similar to \tinycfa.
%Its core idea is to log (and send to \vrf) all the inputs received by the program during its execution, in addition to the program's control flow path (as in \CFA).
%With all inputs and the control flow path, \vrf can emulate the entire data flow of the attested program. Doing so allows \vrf to detect both control flow attacks and illegal modifications to the program's data flow. 
%
As shown in Figure~\ref{fig:workflow}, at compile time, \dialed uses \tinycfa for \CFA-related instrumentation. Additionally, it adds its own instrumentation to log all data inputs to a dedicated memory region, called \ilog.
\dialed's instrumenter defines any non-local variables as data inputs, i.e., any value located \emph{outside} of the attested program's current stack. 
%
%The program's current stack is the region located within the current stack pointer value (top of the stack) and
%the value of the stack pointer when the attested program was first called (base of the program's stack). 
%It includes all local variables. 

Following this definition, any instructions that access data from arguments, peripherals, network, or general-purpose I/O are considered data inputs and recorded to \ilog.
Conversely, reads occurring during regular computation, e.g., instructions that make use of local variables are excluded from \ilog, as they are not inputs to this program. This approach helps keep the size of \ilog relatively small.
	
Recall that \tinycfa instruments the executable to produce \cflog.
In the context of \dialed, both \cflog and \ilog are included in \apex's authenticated output region $OR$. 
As $OR$ is covered by the IEF, \vrf is assured of the integrity of these logs. 
With the code, its execution's control flow path, and all inputs, \vrf can locally emulate execution and its data flow. Therefore, it can verify all steps in this computation, and detect data-only and control flow attacks.

As illustrated in Figure~\ref{fig:arch} and similar to \tinycfa, \dialed requires no hardware support other than \pox. \dialed's instrumentation overhead includes logging inputs (which are typically small in number). Similarly, \ilog size depends on the number of arguments the application receives when it is invoked and the inputs it processes during its execution.

\section{A Comparison of Software Integrity Verification Methods}\label{sec:comparison}

\begin{table}[]
\centering
\caption{Qualitative comparison}%: services offered by different architectures for remote software integrity verification}
\label{table:functionality}
\resizebox{\columnwidth}{!}{%
\begin{tabular}{|c|c|c|c|c|c|c|}
\hline
Scheme                                                                     & \begin{tabular}[c]{@{}c@{}}Baseline\\ MCU\end{tabular} & \begin{tabular}[c]{@{}c@{}}\vrased\\ \cite{vrasedp}\end{tabular} & \begin{tabular}[c]{@{}c@{}}\apex\\ \cite{apex}\end{tabular} & \begin{tabular}[c]{@{}c@{}}\tinycfa\\ \cite{tinycfa}\end{tabular} & \begin{tabular}[c]{@{}c@{}}\dialed\\ \cite{dialed}\end{tabular} & \begin{tabular}[c]{@{}c@{}}\acro\\ \cite{rata}\end{tabular} \\ \hline\hline
\begin{tabular}[c]{@{}c@{}}Detection of\\ Modified Code\end{tabular}       & \xmark                                                 & \cmark                                                    & \cmark                                                  & \cmark                                                     & \cmark                                                    & \cmark                                                  \\ \hline
\begin{tabular}[c]{@{}c@{}}Provable\\ Execution\end{tabular}               & \xmark                                                 & \xmark                                                    & \cmark                                                  & \cmark                                                     & \cmark                                                    & \xmark                                                  \\ \hline
\begin{tabular}[c]{@{}c@{}}Detection of Control\\ Flow Attack\end{tabular} & \xmark                                                 & \xmark                                                    & \xmark                                                  & \cmark                                                     & \cmark                                                    & \xmark                                                  \\ \hline
\begin{tabular}[c]{@{}c@{}}Detection of Data\\ Flow Attack\end{tabular}    & \xmark                                                 & \xmark                                                    & \xmark                                                  & \xmark                                                     & \cmark                                                    & \xmark                                                  \\ \hline
\begin{tabular}[c]{@{}c@{}}TOCTOU\\ Security\end{tabular}                  & \xmark                                                 & \xmark                                                    & \xmark                                                  & \xmark                                                     & \xmark                                                    & \cmark                                                  \\ \hline
\end{tabular}%
}
\end{table}
%
% \begin{figure*}[hbtp]
% 		\centering
% 		\subfigure[LUT]
% 		{\includegraphics[width=0.5\columnwidth]{graphics_python/lut_comparison.pdf}\label{fig:lut}}
% 		\hfill
% 		\subfigure[Register]
% 		{\includegraphics[width=0.5\columnwidth]{graphics_python/reg_comparison.pdf}\label{fig:regs}}
% 		\hfill
% 		\subfigure[Code size]
% 		{\includegraphics[width=0.5\columnwidth]{graphics_python/code_size_comparison.pdf}\label{fig:code_size_comp}}
%         \hfill
% 		\subfigure[Runtime]
% 		{\includegraphics[width=0.5\columnwidth]{graphics_python/runtime_comparison.pdf}\label{fig:runtime_comp}}
% 		\caption{Hardware and software overhead of different architectures.}\label{fig:overhead}
% 		\vspace{-0.2cm}
% \end{figure*}

\begin{figure*}[hbtp]
		\centering
		\subfigure[LUT and Registers]
		{\includegraphics[width=0.75\columnwidth]{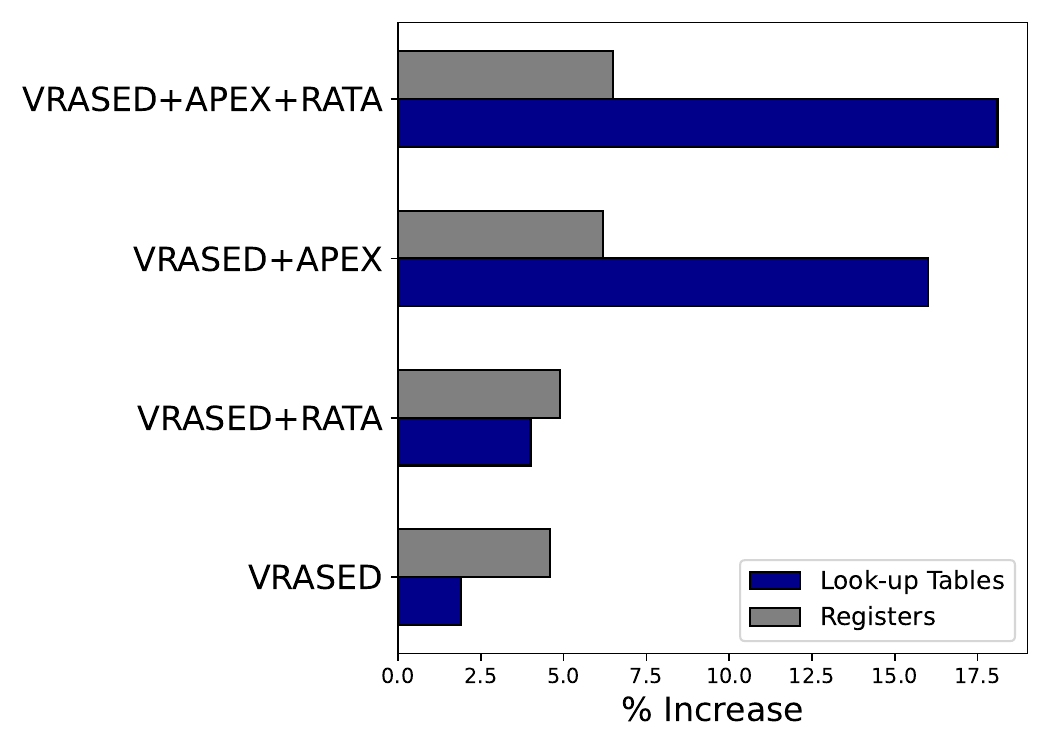}\label{fig:hw_comp}}
		\hfill
		\subfigure[Code Size and Runtime]
		{\includegraphics[width=1.25\columnwidth]{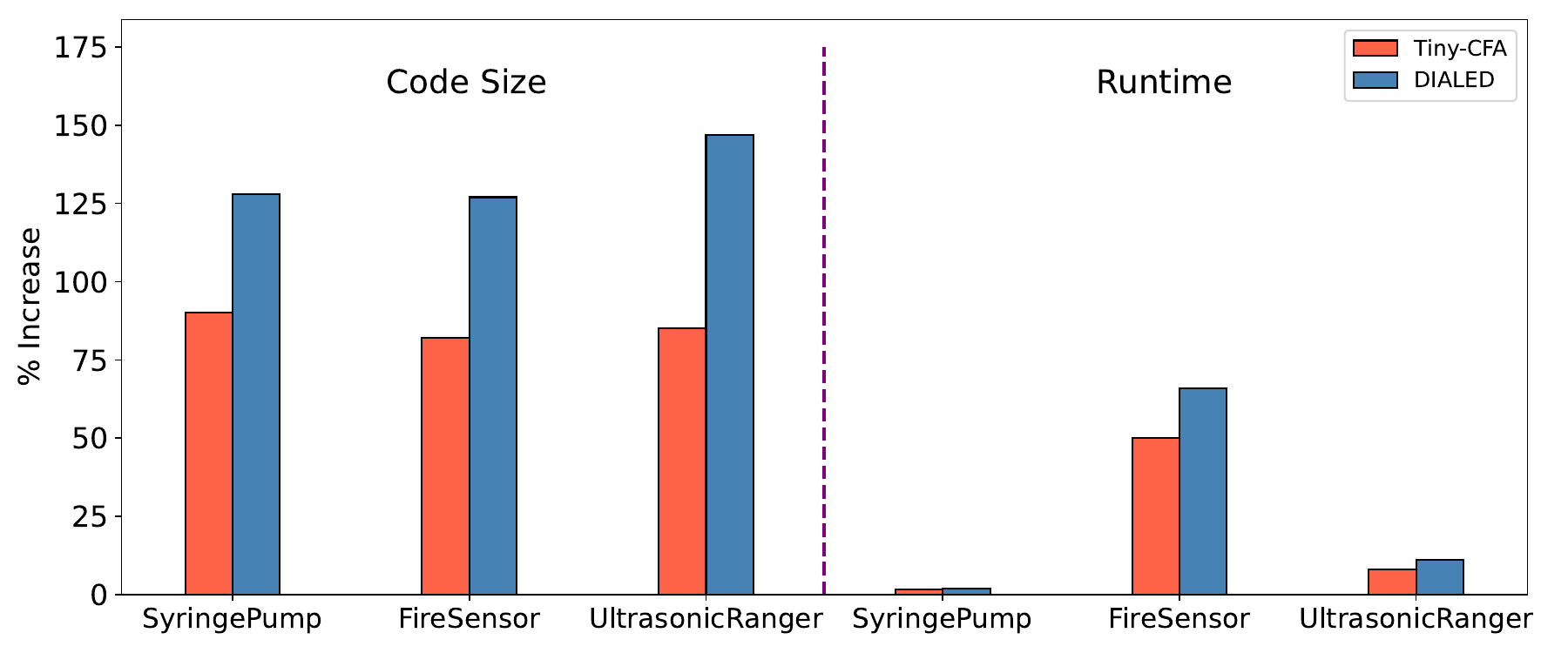}\label{fig:sw_comp}}
		\caption{Hardware and software overhead of different architectures.}\label{fig:overhead}
		\vspace{-0.2cm}
\end{figure*}

This section compares the architectures discussed thus far.
% in both qualitative and quantitative terms. 
Table~\ref{table:functionality} presents a qualitative comparison, highlighting the type of security service offered by each architecture.
Meanwhile, Figure \ref{fig:overhead} reports a quantitative comparison, depicting hardware and software overheads.

\add{As we target low-cost/low-power MCUs, we compare these architectures by instantiating them on OpenMSP430, an open-source version of the TI MSP430 MCUs.}
\remove{These architectures have been implemented on top of OpenMSP430, an open-source version of the TI MSP430 MCU.}
Therefore, the unmodified MSP430 is used as a reference baseline for comparison. 
\add{Although our evaluation is conducted on MSP430, we emphasize that the suitability of these architectures extends to other MCUs within the same class (recall Section~\ref{sec:scope}), e.g., AVR ATMega and ARM Cortex-M MCUs. However, the proprietary nature of these designs precludes direct evaluation.}
Following the common practice in this space~\cite{smart,dessouky2018litehax}, %
the hardware overhead is reported in terms of additional Look-Up Tables (LUTs) and registers. Added LUTs represent increase in combinatorial logic, whereas added registers are due to sequential logic required in each case. 

Figure \ref{fig:hw_comp} depicts percentage increase in hardware relative to the CPU core cost. \vrased hybrid \RA architecture offers the simplest type of integrity evidence, i.e., whether \prv is currently loaded with the correct software image. Atop the baseline MCU, it incurs around $2$\% additional LUTs and 4.5\% additional registers. 
\apex, which is a superset of \vrased's hardware support, offers both \RA and \pox and requires 16\% extra LUTs and 6\% extra registers. 

\tinycfa and \dialed add support for \CFA and \DFA.
Since they rely on \apex hardware support for \pox ``as is'', they do not incur additional hardware costs.
However, they lead to notable code size and runtime overhead as a result of the instrumentation phase. 
While the exact overhead is application-dependent (due to the variable number of branch instructions
in different applications), we report \tinycfa and \dialed overheads on three real-world open-source MCU applications: Open Syringe Pump, Fire Sensor, and Ultrasonic Ranger. Details of these applications can be found in~\cite{dialed}.

Figure \ref{fig:sw_comp} shows the percentage code size and runtime increase caused by instrumentation. On average, \tinycfa increases the code size by 85\% and \dialed by 130\%. 
Whereas, the runtime increase is approximately between $2 - 65$\% over their non-instrumented counterparts.

\acro can be viewed as an add-on to any of the aforementioned architectures to provide \toctou-security and reduced \RA computation time. When added to \vrased alone it adds 4\% LUTs and 6\% registers.

When considered in conjunction (as depicted in Figure~\ref{fig:arch}), all discussed features add up to 18\% LUTs and 7\% registers to the baseline MCU. Together they provide \toctou-Secure and faster \RA, \pox, as well as hardware support required by instrumentation-based \CFA and \DFA. 

\section{Open Problems and Opportunities}\label{sec:opportunities}

This section outlines open challenges and future research directions in this area.

{\it 1) Formal Verification \& Provable Security:}
%The architectures discussed in this article aim to produce unforgeable evidence for the software state of their host MCU. 
The trustworthiness of software integrity proofs also heavily depends on the correct implementation of the underlying architectures.
\emph{Formal verification} is a common approach to prove the correctness of the system implementation with respect to formal design specifications. \vrased, \apex, and \acro already employed this approach to ensure security/correctness in their hardware and software implementations.
Nonetheless, there are no formally verified architectures for \CFA or \DFA.
\remove{In the context of \tinycfa and \dialed, this entails formally verifying the instrumentation phase, which can be an interesting research direction.}
\add{The main challenge in verifying Tiny-CFA and DIALED lies in how to verify security and correctness of the instrumentation phase. Given the absence of this phase in VRASED/APEX/RATA, the verification methods employed by these architectures cannot be directly applied here. 
Addressing this presents an interesting avenue for future research.}
%We are unaware of any prior work tackling this issue and believe this could be one of the interesting research directions in this area.
	
{\it 2) Higher-End Devices:}
This article examines software integrity techniques in resource-constrained MCUs.
One avenue for future research involves extending these guarantees to higher-end devices.
General-purpose CPUs (e.g., those featured in smartphones or desktops) are not as cost-prohibitive and thus often come equipped with more sophisticated hardware (e.g., MMUs or Trusted
Execution Environments) or software modules (e.g., micro-kernels or monolithic operating systems).
A promising opportunity for future research is to leverage the added hardware and software support to obtain similar guarantees to those considered in this article.
%and (ii) how to obtain these guarantees provably perhaps by leveraging formal verification at the protocol and implementation levels. 
	
\remove{
{\it 3) Confidentiality \& Availability:}
While the article focuses on software integrity problems, 
confidentiality and availability are also important concerns for IoT devices.
Given their prevalence in sensitive data collection applications (e.g., recording pulse rate in medical devices), the absence of confidentiality protection would render such sensitive data vulnerable to unauthorized access by \adv.
Moreover, these devices are increasingly deployed in safety-critical settings.
Without availability guarantees, \adv may prevent \prv from completing its safety-critical task (e.g., sounding an alarm upon detecting a gas leak).
While software integrity is required for achieving confidentiality and availability, it is not sufficient in itself. Future research should consider addressing confidentiality and availability issues in IoT devices.}

{\it 3) Efficiency of \CFA and \DFA:}
While \tinycfa and \dialed provide relatively low-cost \CFA and \DFA, the code size and run-time increases are still significant. Furthermore, as attested operations increase in size and complexity, the generated evidence traces (\cflog and \ilog) also increase accordingly. As a consequence, \CFA and \DFA are still limited to simple self-contained operations in which associated evidence can be stored and transmitted by a resource-constrained MCU. An interesting direction for future work lies in the management and reduction of \CFA and \DFA associated costs.

\ignore{
{\it 5) Interpreting \CFA and \DFA Evidence:}
Even when implemented correctly, the techniques discussed in this article are only useful if the \vrf can efficiently analyze the reported evidence to decide if a given report represents a valid state/execution.
In \RA, this can be achieved via a simple equality check to the expected software image. However, when analyzing control flow and data flow evidence, this problem becomes non-trivial and has been, for the most part, overlooked in the literature.
As the space of possible control flows and data flows is typically infinite, 
%(even for simple programs/tasks), 
future work must look into how to employ software analysis methods %, including static/dynamic analysis, 
to automate the verification of \CFA and \DFA evidence.
}

{\it 4) Attesting vs. Auditing Software Integrity:}
Current architectures enable only \textit{detection} of software compromises.
They cannot guarantee that \vrf ever receives the produced evidence, in case of software attacks. While this suffices to detect if the \prv is compromised in a yes/no manner (in general, the absence of a signed report from the \prv indicates that \underline{something is wrong}), 
it precludes {\it auditing} the generated evidence to pinpoint the source of compromises 
(i.e., to determine \underline{what is wrong} with the \prv's software).
Auditing is non-trivial because a compromised \prv might ignore the protocol and simply refuse to send back evidence that indicates a compromise. Resolving this issue remains an open problem for future work.

\add{5) Multi-Device Settings: This article focuses of software integrity techniques in a single-\dev setting. 
However, many IoT systems rely on a large group (``swarm'') of interconnected devices.
Simply applying single-\dev solutions to the swarm setting faces scalability issues.
To address this, in the context of \RA, several ``swarm/collective \RA'' techniques have been proposed to efficiently perform RA across a multitude of devices. These techniques vary in their target settings, considering different factors, e.g., swarm topologies, swarm dynamics, and software/hardware heterogeneity. We refer to~\cite{ambrosin2020collective} for an in-depth overview of swarm \RA.}
\add{Notably, \acro holds the potential to enhance existing swarm attestation schemes. With its advantages of constant runtime and TOCTOU security, RATA can serve as a building block in swarm \RA schemes, yielding faster overall swarm attestation while ensuring synchronized TOCTOU security across the swarm. 
Although swarm \RA has been extensively studied, the extension of other integrity services, i.e., PoX/CFA/DFA, to a swarm of devices remains largely unexplored, presenting an opportunity for future work.
}

\section{Conclusions}
This article overviews a series of techniques for remote verification of software integrity on 
resource-constrained IoT devices. 
Service provided by each technique varies, starting from simply verifying code installed 
on a remote device, to detection of transient malware, and detection of runtime attacks that 
arise due to vulnerabilities in installed code.
%, such as control flow and data-only attacks. 
Not surprisingly, techniques that provide more sophisticated service are also accompanied by increased complexity 
and costs. To assess these trade-offs, we compare them qualitatively and quantitatively, considering the security 
services provided in each case, additional hardware cost, and runtime overhead.
Finally, we present new research directions and open problems. 

\bibliographystyle{ieeetr}
\bibliography{references}
\vspace{-0.5em}
\section*{Biographies}

\noindent Ivan De Oliveira Nunes (ivanoliv@mail.rit.edu) is an Assistant Professor at the Rochester Institute of Technology (RIT). Before RIT, he received a Ph.D. from the University of California Irvine. His research interests span the fields of Security \& Privacy, Computer Networking, Computing Systems, and especially their intersection.
~\vspace{2mm}

\noindent Sashidhar Jakkamsetti (sashidhar.jakkamsetti@us.bosch.com) is a Research Scientist at Robert Bosch LLC - Research and Technology Center. He obtained his Ph.D. from the University of California, Irvine.
%and a Bachelor's degree from the Indian Institute of Technology, Roorkee (2016). Previously, he worked as a Software Engineer at Microsoft in India (2016-2018). 
His research focuses on IoT Security/Privacy, Applied Cryptography, and Privacy-Preserving Technologies.
~\vspace{2mm}

\noindent Norrathep Rattanavipanon (norrathep.r@phuket.psu.ac.th) received a Ph.D. from the University of California, Irvine. He is an Assitant Professor with the College of Computing, Prince of Songkla University, Phuket Campus. 
His research interests include IoT security as well as software and binary analysis.~\vspace{2mm}

\noindent Gene Tsudik (gene.tsudik@uci.edu) received all his degrees eons ago. He's a fellow of all the usual societies. He dabbles in numerous security/privacy and applied crypto topics. He also occasionally composes atrocious crypto-poetry. 

\end{document}